\documentclass[12pt]{article}
		\usepackage{psfig,cite}
\begin{document}
\normalbaselineskip=16 true pt
\normalbaselines
\bibliographystyle{unsrt}

\def\be {\begin{equation}}
\def\ee {\end{equation}}
\def\lr {\longrightarrow}
\def\dis {\displaystyle}
\def\gev{\: {\rm GeV} }
\setcounter{page}{0}
\thispagestyle{empty}
\renewcommand{\thefootnote}{\fnsymbol{footnote}}

\begin{flushright}
MRI-PHY/P980342\\[1.5ex]
{\large \tt hep-ph/9804297}
\end{flushright}

\vskip 1 true cm

\begin{center}

{\Large\bf Observing Doubly Charged Higgs Bosons \\[3ex]
           in Photon-Photon Collisions
}\\[10mm]
{\large\em 
Surajit Chakrabarti$^{a,}$\footnote{surajit@cubmb.ernet.in}, 
Debajyoti Choudhury$^{b,}$\footnote{debchou@mri.ernet.in}, 
Rohini M. Godbole$^{c,}$\footnote{rohini@cts.iisc.ernet.in; 
			On leave of absence from University of Bombay, India}
and Biswarup Mukhopadhyaya$^{d,}$\footnote{biswarup@mri.ernet.in; On leave
            from Mehta Research Institute, Allahabad - 211 019, India}
}
\end{center}

\begin{abstract}

We discuss the possibility of observing doubly charged
Higgs bosons in $\gamma \gamma$ collisions. We find that one can 
increase the range of observability close to the kinematic limit
by a judicious choice of the polarisations of the initial $e^-/ e^+$
beams as well as the initial laser beam photons which are made 
to back-scatter from the former.
We also note that, in a large region of parameter space, the generally 
used lepton number violating
decay mode is dominated by the decay into a singly charged Higgs and a 
real or virtual $W$-boson, giving rise to a large number of 
fermions in the final state. This can qualitatively alter the strategy for
discovering doubly-charged scalars.

\end{abstract}

\vskip 1 true cm

\begin{center}
{\it
$^a$Physics Department, Maharaja Manindra Chandra College, 
Calcutta - 700 006, India\\
$^b$Mehta Research Institute of Mathematics and Mathematical Physics,
	  Chhatnag Road, Jhusi, Allahabad - 211 019, India\\
$^c$Centre for Theoretical Studies, Indian Institute of Science, 
Bangalore - 560 012, India\\
$^d$Theory Division, CERN, CH 1211 Geneva 23, Switzerland\\
}
\end{center}

\setcounter{footnote}{0}
\renewcommand{\thefootnote}{\arabic{footnote}}
\setcounter{page}{0}
\pagestyle{plain}
\advance \parskip by 10pt
\pagenumbering{arabic}

\newpage

\section{INTRODUCTION}

The symmetry breaking sector of the electroweak theory is yet to be
confirmed experimentally. The standard one Higgs-doublet scenario
is by no means the only, though certainly the simplest, option. As
a matter of fact, an extended Higgs sector appears quite naturally in 
many options beyond the Standard Model(SM), which have been suggested 
to  handle some of the theoretical problems posed by the SM iteself.
Hence models with a richer scalar sector are  constantly
under investigation. A further motivating factor in these studies is  
the fact that the up-and-coming
experiments are capable of testing many of the  extensions
of the Higgs sector. Of course, an immediate generalisation of the 
Glashow-Salam-Weinberg scheme is to introduce additional
scalar doublet(s), leading to singly- charged physical scalar(s). 
A further step is to consider the possibility 
of doubly charged scalar bosons, and the
phenomenology entailed by them.

Doubly charged scalars occur in many extensions of the standard 
model~\cite{hhunter}. They usually come as members of triplet Higgs bosons, 
the most common examples of such models being those with left-right 
symmetry~\cite{moh}.  All the models containing Higgs triplets 
have the merit of being able to provide lepton-number violating Yukawa
couplings, thereby enabling one to introduce left-handed Majorana masses
for neutrinos~\cite{moh,neutr}. 

It is needless to say that the phenomenology of these exotic Higgs 
bosons~\cite{gunion} is quite interesting. A number of 
new channels open up when they decay. Also, these scalars may be endowed with
lepton number violating couplings, leading to additional physics 
possibilities. The signals for the singly charged Higgs bosons
occuring in these models at the current and future $e^+e^-$
colliders have  been studied in some detail~\cite{us1,us2,apostolos}.
Constraints on the lepton flavour violating couplings of 
the doubly charged Higgs bosons possible from low energy
processes as well as the $e^+ e^-$ experiments at 
PEP/PETRA~\cite{swartz, picciotto} as well as the region that can be
explored at future linear colliders~\cite{raidal} have also been
discussed in the literature. The search possibilities for
these at the $e^+e^-/e^-e^-$ colliders~\cite{lusignoli,cuypers}
as well as at the hadronic colliders~\cite{loomis,raidal2,ackr}
have been investigated to some extent.
In this paper we want to point out that photon-photon
collisions~\cite{fawzi}, implemented through laser 
back-scattering in $e^{+}e^{-}$ machines, are particularly suitable for
studying doubly charged  Higgs bosons. This is because the production
mechanism is model-independent and depends entirely upon  electromagnetic 
interaction. The production rate is enhanced due to the `double'  
electric charge of the pair-produced scalars. The recent observation
of {\it photon photon} scattering using backscattered 
lasers~\cite{slac} has raised the hope that 
$\gamma \gamma$ colliders~\cite{telnov} can be made to work.

After studying the polarisation dependence of the total production
cross-section, we go on to investigate the decays of such scalars in the
most common forms of models. Our conclusion is that though like-sign
dileptons are frequently referred to as the most striking signals of a
doubly-charged Higgs, other channels are often of greater importance.  
This is particularly noteworthy in view of the fact that the $\Delta L = 2$
(dilepton) channel is allowed {\it only in a triplet Higgs scenario}.
If for some reason a doubly charged scalar scalar occurs (together with a
singly charged companion) in a higher representation of $SU(2)$, then the 
dileptons will not arise, whereas other decays will still be possible,
 most important among them
being $H^{++} \longrightarrow H^{+} W^{+}$ driven by SU(2) gauge coupling.  
Our purpose is to draw attention to these other channels in the 
context of search strategies adopted for the 
doubly charged Higgs scalars in future experiments.

In section 2 we briefly outline the framework within which we operate in this
study, and the constraints on the various parameters in this framework.
Section 3 reports on the calculated cross-sections for the pair-production
of doubly-charged Higgs bosons is $\gamma - \gamma$ collision.
A study of the various possible decay channels is taken up in section 4.
Our conclusions are summarised in section 5.

\section{THE SCENARIO AND THE PARAMETERS}

There are many theoretical models which 
necessitate the existence of doubly charged Higgs bosons. Here we assume
only the most common feature of such models, namely, that the doubly-charged
scalar occurs in an exotic representation of SU(2) (a triplet in most cases).
The usual SU(2) doublet Higgs mixes with this exotic multiplet through the
scalar potential, the mixing being parametrized by an angle
$\theta_H$. Furthermore, if the exotic multiplet is a triplet, it can have
lepton-number violating couplings. We accommodate this feature also in our 
scenario, and parametrize the strengh of the $\Delta L = 2$ coupling by
$h_{ll}$. 

  When $\theta_H$ is specified, one automatically fixes the ratio of the
vacuum expectation values (VEV) of the triplet and the doublet:
\be
s_H = {\frac{\sqrt{2} w}{\sqrt{v^2 + 2 w^2}}}
\ee
where $v$ and $w$ are respectively the VEV's of the doublet and 
the triplet, and $s_H = Sin \theta_H$. 
  
 An immediate consequence of a non-zero VEV of the triplet is 
a contribution to the $\rho$-parameter, defined as 
$\rho = m^2_W/{m^2_Z Cos^{2}{\theta_W}}$. With $x = w/v$, 
\be
\rho = {\frac{1 + 2x^2}{1 + 4x^2}}
\ee

The experimental constraint on $\rho$ thus severely restricts
the value of $\omega$ and, in turn, $\theta_H$. It can be seen that the current
data lead to the limit $s_H \leq 0.0056$ at 95\% confidence 
level~\cite{pdg}.

The couplings $h_{ll}$ ($l = e, \mu \tau$)  are also subject 
to constraints from
lepton-number violating processes. These include the contribution of
$\Delta L = 2$  vertices to Bhabha scattering, anomalous magnetic 
moment of the muon, muonium-antimuonium conversion, the decay 
$K^{+} \lr \pi^{-} \mu^{+} \mu^{+}$ etc~\cite{swartz,picciotto,
raidal}. But by far the strongest limits
are obtained using the upper bounds on the (Majorana) masses of the three
neutrinos implying that ${h_{\tau \tau} \leq 1.4 \times 10^{-4}}/s_H$. The 
couplings for the first two generations have to be down by 3 and 8 orders
respectively compared to this value. Essentially, this means  that the 
lepton-number violating couplings of the doubly charged Higgs can perhaps be
phenomenologically significant for the third generation only. Also, the smaller
is $s_H$, the larger becomes the possibility of having a large $\Delta L = 2$
interaction.

Finally, there are the masses of the doubly-and singly charged scalars
($m_{++}, m_{+}$), which we use as free parameters in  the present study. 
The only other parameter which is potentially relevant for the our 
calculations is the self-coupling involved in the $H^{++} H^{-} H^{-}$ 
interaction. Although this coupling is determined once the masses 
and the triplet VEV are specified, we choose to emphasize our points
by going to a region of the parameter space where the decay 
$H^{++} \lr H^{+} H^{+}$ is not kinematically allowed. This eliminates the
need of using the self-coupling here.
  
Before we end this section, two comments are in order. First, 
the introduction of
Higgs triplets (or any multiplets higher than doublets) in general 
implies the existence of a tree-level interaction involving the W, 
the Z and  a charged physical scalar, a feature not present in any 
scenario containing only doublets. Such an interaction is proportional
to the parameter $s_H$, and is often important in the decays of the
singly-charged component $H^{+}$~\cite{us1,us2}. Side by side, the upper 
limit on the lepton number violating coupling is proportional
to the inverse of $s_H$. This shows a complementarity between the signals
that can be driven by either of these two types of interactions, 
a feature that
should be kept in mind when the signals for such scenarios are 
being explored.

Secondly, there are some models where the constraint on $s_H$ 
from $\rho$ is avoided by postulating complex and real triplets 
together~\cite{manchek,golden}, and assuming a custodial symmetry 
protecting the equality of the VEV's. We are not assuming anything 
of that kind in this paper.

\section{PRODUCTION IN PHOTON-PHOTON COLLISIONS}

As has been already stated above, $\gamma\gamma$-collsions provide a 
model-independent, `democratic' way of producing doubly charged Higgs bosons,
with the advantage that the cross-section is going to be 16 times larger than 
a corresponding case with a singly charged scalar. 

To understand the features of this production mechanism, let us first 
consider a hypothetical $\gamma\gamma$ machine operating
at a {\em fixed center-of-mass energy} $\sqrt{s}$. 
The expression for the differential cross-section is given by 
\be 
\begin{array}{rcl}
\dis
\frac{ {\rm d} \sigma}{ {\rm d} \Omega} 
	& =  & \dis
          \frac{16 \alpha^2}{ s} \; \beta \;
           \Bigg[  \frac{ 1 + \xi \cdot \chi}{2} 
		 - \; \frac{ \beta^2 s_\theta^2}{1 - \beta^2 c_\theta^2}
			\left\{  1 + \xi \cdot \chi 
				+ (\xi_3 + \chi_3) c_{2\phi}
				+ (\xi_1 + \chi_1) s_{2\phi}
                        \right\}
              \\[2ex]
        & & \dis \hspace*{4em}
		 + \; \left(
			\frac{ \beta^2 s_\theta^2}{1 - \beta^2 c_\theta^2}
                   \right)^2
			\;
			\left( 1 + \xi_3 c_{2\phi} + \xi_1 s_{2\phi}
			\right)
			\;
			\left( 1 + \chi_3 c_{2\phi}  + \chi_1 s_{2\phi}
			\right)
		\Bigg]
\end{array}
	\label{cs:fixed_ener}
\ee
where $s_\psi \equiv \sin \psi$, $c_\psi \equiv \cos \psi$
and $\beta \equiv (1 - 4 m^2 / s)^{1/2}$ is the velocity of the $H^{++}$ 
in the center-of-mass frame. In eq.(\ref{cs:fixed_ener})
$ \xi \equiv ( \xi_1, \xi_2, \xi_3) $ are the 
Stokes parameters for one of the photons 
with $\xi_2$ denoting the circular polarization (helicity). 
Similarly, $\chi_i$ are the Stokes parameters for the other 
photon. The cross-section for unpolarized scattering are obtained 
from eq.(\ref{cs:fixed_ener}) by substituting $\chi_i = \xi_i = 0$. 

\begin{figure}[htb]                
        \centerline{
            \psfig{figure=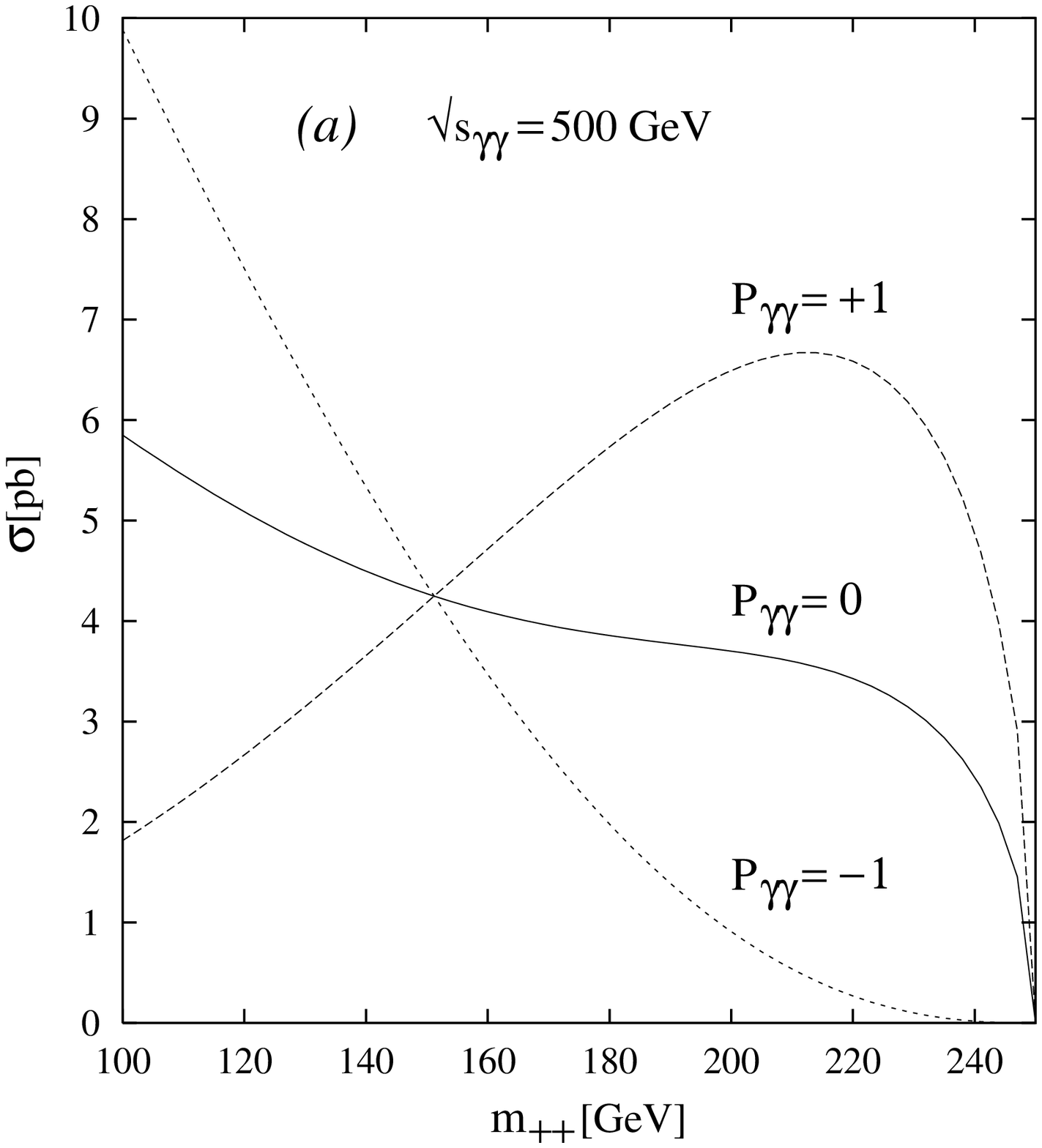,width=8.5cm,height=7cm,angle=0}
            \psfig{figure=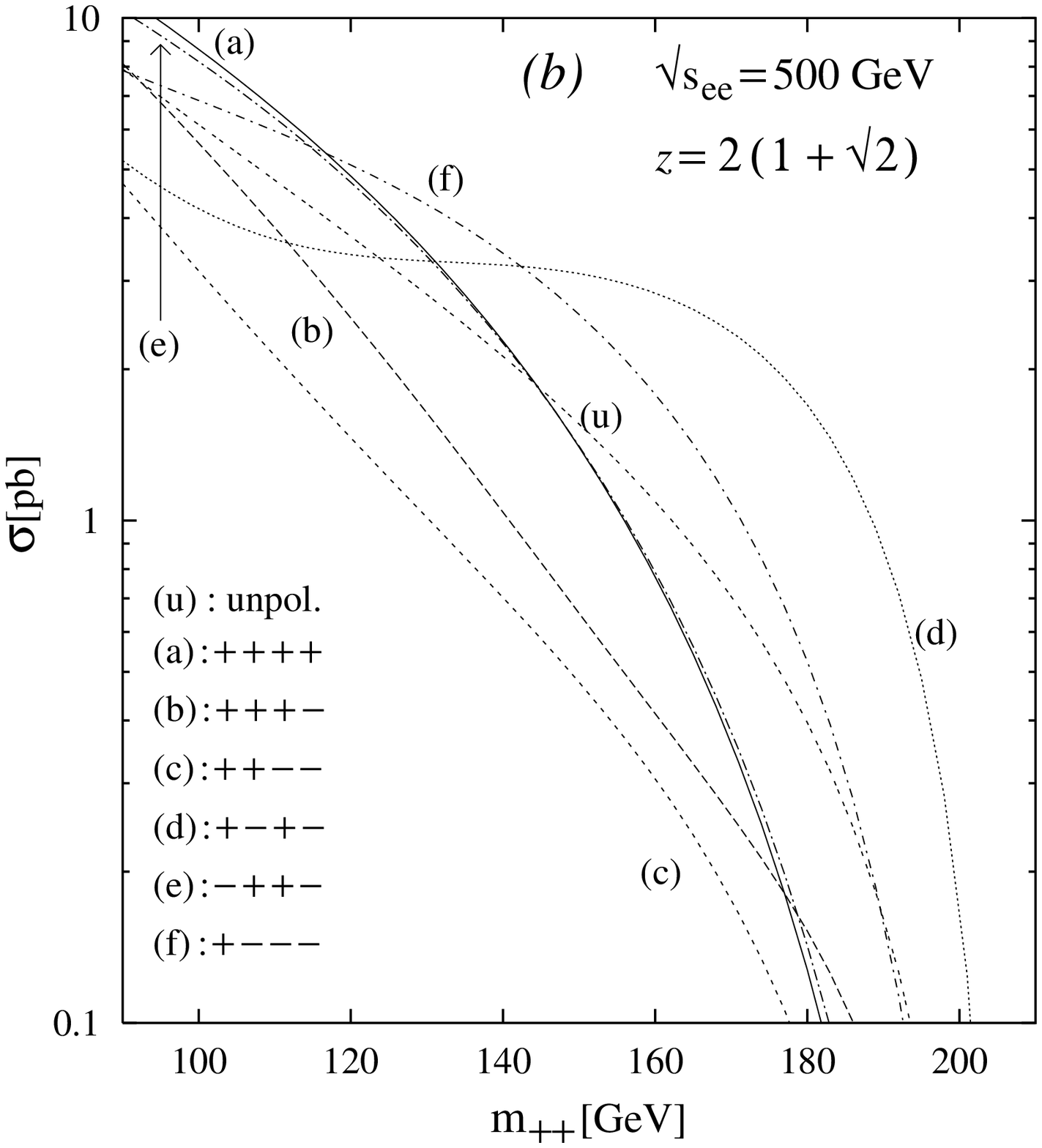,width=8.5cm,height=7cm,angle=0}
                   }
        \caption[sig]{\em Production cross-sections as a function 
			of $m_{++}$. 
		      {\em (a)} For a monochromatic $\gamma \gamma$ machine
				and three choices of the product of 
				photon helicities.
		      {\em (b)} For a realistic 
				(see eq.(\protect\ref{spectrum})) 
				$\gamma \gamma$ collider operating at 
				a fixed $s_{ee}$.  
				The successive polarisation signs 
				in the labels correspond to 
				($e_1, {\rm laser}_1, e_2, {\rm laser}_2$)
				in that order.
			     }
        \label{fig:cs}                                                   
\end{figure}

For the sake of simplicity, let us consider only circularly polarised 
photon beams. Obviously, the differential cross section should 
now possess an azimuthal symmetry and this is reflected by the 
expression above. In fact, all the polarisation dependence 
now appears through the first two terms 
of eq.(\ref{cs:fixed_ener}) and is proportional to the product 
$P_{\gamma \gamma} \equiv \chi_2 \xi_2$. 
In Fig.\ref{fig:cs}($a$), we show the mass-dependence of this 
cross-section for three different values of  this product.
The intersection point of the curves is given by the transcendental 
equation $ (1 + \beta) / (1 - \beta) = exp(2 \beta)$. 
Clearly for a small higgs mass, the $P_{\gamma \gamma} = -1 $ mode 
is preferable, while close to the kinematical limit, the 
$P_{\gamma \gamma} = 1 $ mode wins easily. This feature needs to be 
remembered in designing the experiment.

In practice, however, the only known way to achive high energy $\gamma \gamma$ 
collisions is to affect laser back-scattering 
in a linear $e^{+}e^{-}$ or $e^{-}e^{-}$ collider~\cite{telnov}. 
The spectrum of the Compton-scattered photons is determined 
by the polarization of the electron and the laser and one combination
of their energies and the incident angle, namely,
\[
	z \equiv \frac{4 E_b E_\ell}{m_e^2} 
                 \cos^2 \frac{\theta_{b \ell}}{2}
	\ .
\]
The quantity $z$ also determines the maximum value of the fraction $y$ 
of the electron energy $E_e$ carried by the photon ($y = \omega/E_b$)
\[
y_{\rm max} = \frac{z}{1 + z}
\]
It may seem that $z$ and hence $\omega_{max}$ can 
be enhanced by increasing the energy of the incident laser beam. However, 
an arbitrary increase in that energy makes the process inefficient because
of electron-positron pair production through interaction  of the incident and  
scattered photons. An optimal choice in this respect is 
$z = 2 (1 + \sqrt{2})$, which we have adopted in our calculation.

If the initial laser photons are circularly polarised ($P_\ell$)
and the electron polarisation is $P_e$, then the photon number 
density and average helicity are given by~\cite{telnov}
\be
\begin{array}{rcl}
\dis \frac{ {\rm d} n} { {\rm d} y } 
	& = & \dis \frac{2 \pi \alpha^2}{m_e^2 z \sigma_C} \; C(y)
   \\[3ex]
\xi_2 (y) & = & \dis 
	\frac{1}{C(y)} \left[ P_e \left\{ \frac{y}{1 - y} + y (2 r - 1)^2
				  \right\}
			     - P_\ell (2 r - 1) \left( 1 - y + \frac{1}{1 -y}
						\right)
		       \right]
   \\[3ex]
C(y) & = & \dis
	\frac{1}{1-y} + (1-y) - 4r(1-r) 
         - 2 P_e P_\ell r z (2r-1)(2-y)
\end{array}
	\label{spectrum}
\ee
In eq.(\ref{spectrum}), $\sigma_C$ is the total Compton cross-section 
and $r \equiv y / z / (1 - y)$. 

The actual cross-sections can then be calculated by folding the 
cross-section of eq.(\ref{cs:fixed_ener}) with the above spectrum
for various polarization choices. While it is relatively straightforward
to have a fully polarized laser, it might be difficult to achieve 
more than 95\% polarization for an electron. In the rest of the study,
whenever we mention polarized beams, we shall understand it to mean
\[
	\left| P_\ell \right| = 1,  \qquad \left| P_e \right| = 0.9 
     \ .
\]

In Fig.\ref{fig:cs}($b$), we plot the production rates 
for a collider operating at 
$\sqrt{s_{ee} }$ = 500 GeV. We show results both for unpolarised 
beams and for configurations with different combinations of 
electron and  laser polarisation. Several comments about the cross
sections are in order:
\begin{itemize}
\item The functional behaviour is determined by two sources: the dynamical 
      dependence on the product $\xi_2 \chi_2$ and the shape of the 
      spectrum.
\item The curves remain invariant under simultaneous reversal of all 
      polarizations.
\item The unpolarised case is the incoherent average of the other six.
\item ${\rm d} n /{\rm d} y$ has a  high-energy bump for 
      cases (d) and (e). For the others, it is monotonically 
      decreasing.
\item For cases (a) and (c), ${\rm d} n /{\rm d} y$ are the same, but 
      the product $\xi_2 (y) \chi_2(y)$ has opposite values. The same 
      comments hold for the pairs (b, f) and (d, e).
\item At small $y$, $\xi_2 (y) \chi_2(y)$ is positive and large 
      for all of (a, b, d). At large $y$, the product is small. 
      While it is monotonically decreasing for (a), it has one and 
      two nodes for (b) and (d) respectively.
\item \begin{tabbing}
       $m_{++} < 120 \gev$ : \hspace{6em} \= $(++++)$ gives largest $\sigma$;\\
       $120 \gev < m_{++} < 140 \gev$ :   \> $(+---)$ gives largest $\sigma$;\\
       $m_{++} > 140 \gev$ :              \> $(+-+-)$ gives largest $\sigma$.
	\end{tabbing}
\end{itemize}

Thus we see that by a proper choice of polarisation combinations we can 
actually increase the range in $m_H$ for a given beam energy.  Since this is 
a reflection of the polarisation dependence of production rate for 
a pair of scalars, one can say by using the equivalence theorem that even the 
$W^+W^-$ production cross-section should show some interesting
interplay between the polarisations of the initial state 
$e^-/e^+$ and laser photons and the final state $W$s. 
 
\section{DECAYS OF THE $H^{++}$}
 
The signals of the doubly-charged Higgs will depend crucially on which
of its decay modes are favoured in the corresponding region of 
the parameter space.
In most studies so far, it has been assumed that the $l^{+}l^{+}$-channel 
reigns supreme and that the main signals are the very conspicuous like-sign
dilepton pairs, with the invariant masses of both the pairs peaking at the
same region. However, we would like to point out here that the dileptonic 
decay channel  is by no means always the most favoured one, especially when 
there exists a singly charged scalar which is lighter than the doubly
charged one. 

Assuming  that the doubly-charged scalar, $H^{++}$, can decay into 
the singly-charged one, $H^{+}$, the following decay channels 
are available to $H^{++}$: 
\begin{enumerate}
\item
$H^{++} \longrightarrow l^+ l^+$  

\item
$H^{++} \longrightarrow W(W^*) W(W^*)$

\item
$H^{++} \longrightarrow H^{+} W(W^*)$

\item
$H^{++} \longrightarrow H^{+} H^{+}$
\end{enumerate}
We have already mentioned that the last mode will be neglected in the
study here. Of the remaining three, the level of domination of (1) clearly
depends on the lepton-number violating coupling $h_{ll}$ which in turn has 
a maximum possible value depending on $s_H$. Therefore, the lower 
the value of $s_H$ is, 
the more insignificant we expect process (2) to be, with a 
corresponding increase in the share of (1).         

\begin{figure}[htb]                
        \centerline{
            \psfig{figure=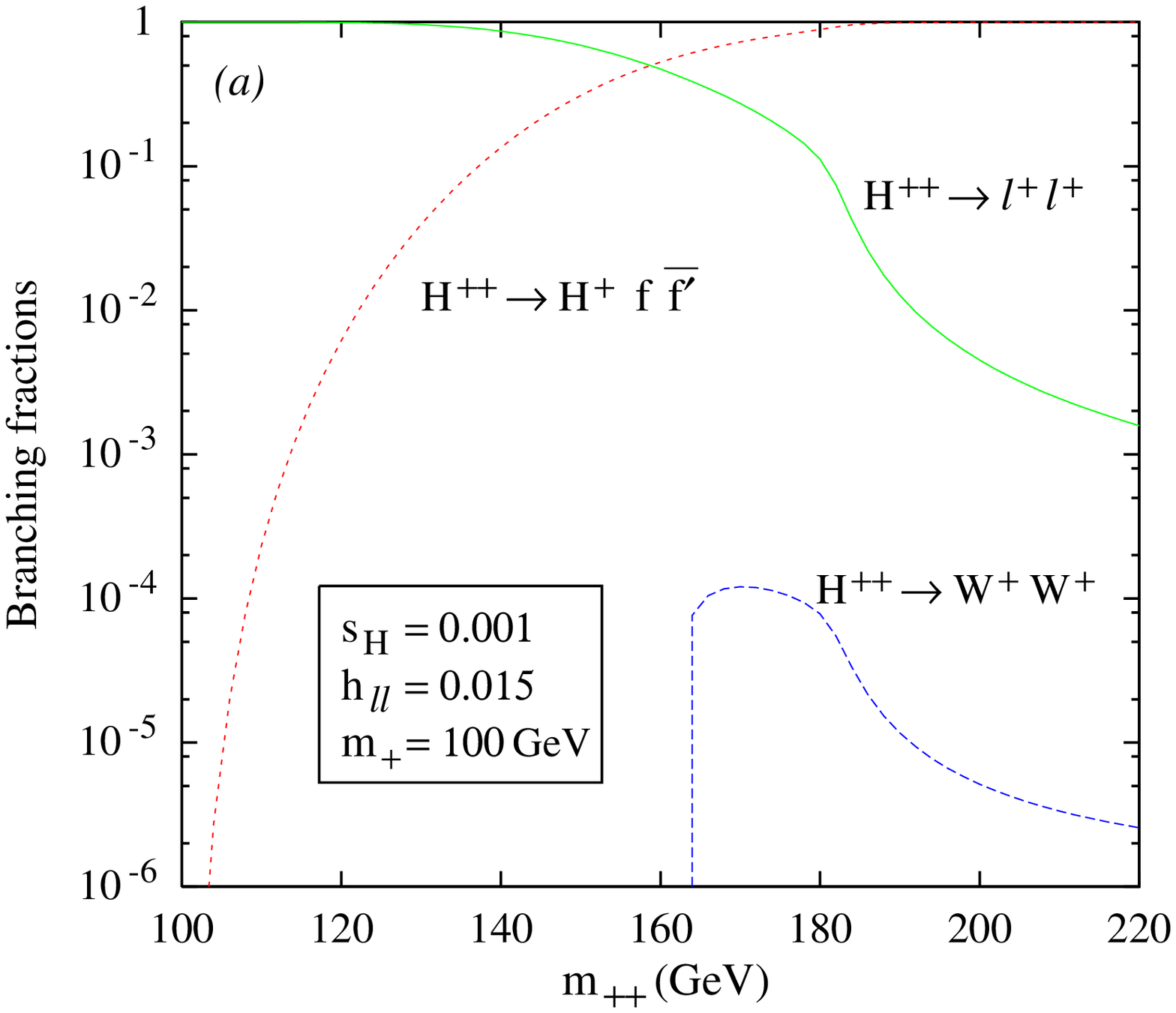,width=8.5cm,height=7cm,angle=0}
            \psfig{figure=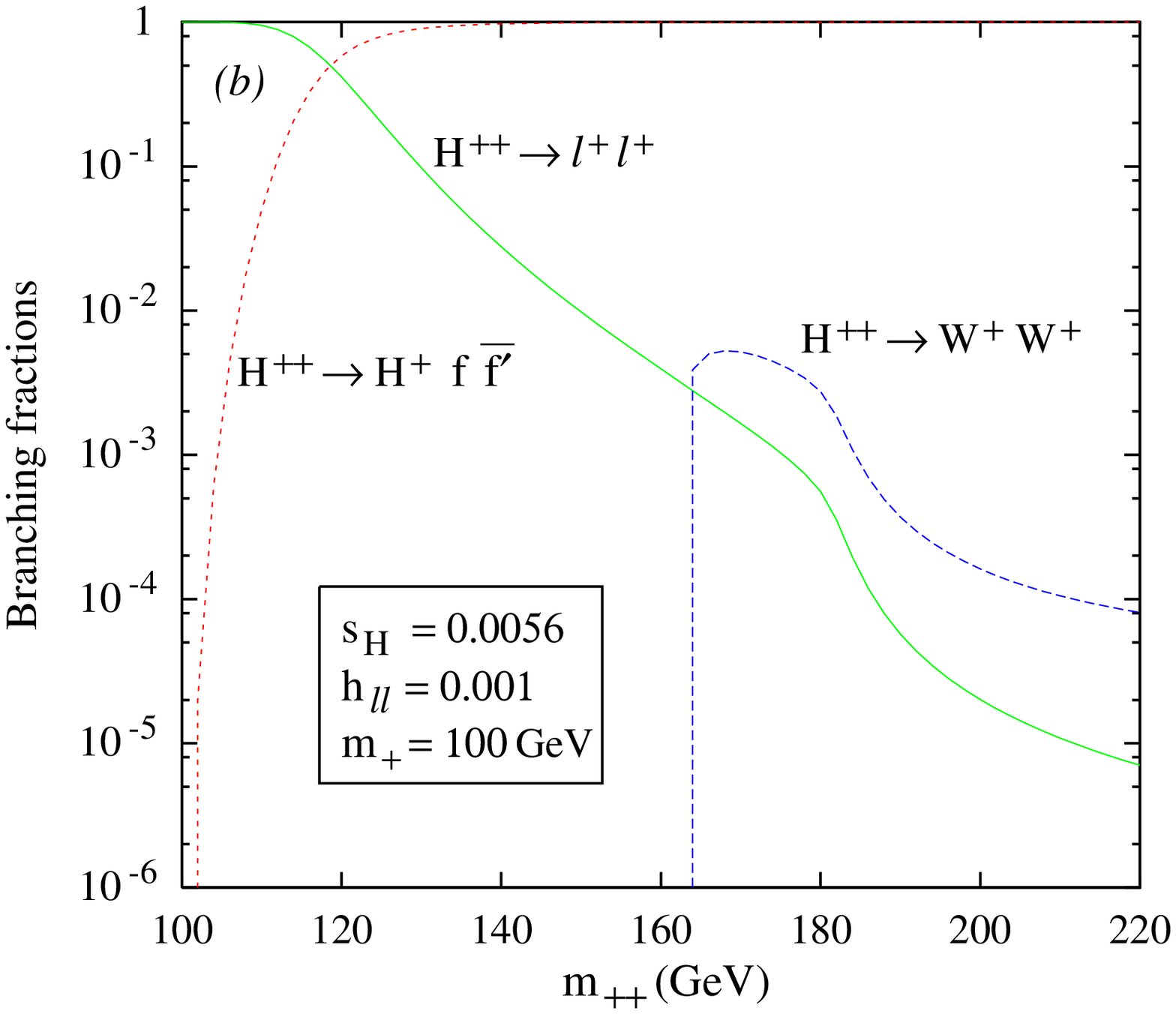,width=8.5cm,height=7cm,angle=0}
                   }
        \centerline{
            \psfig{figure=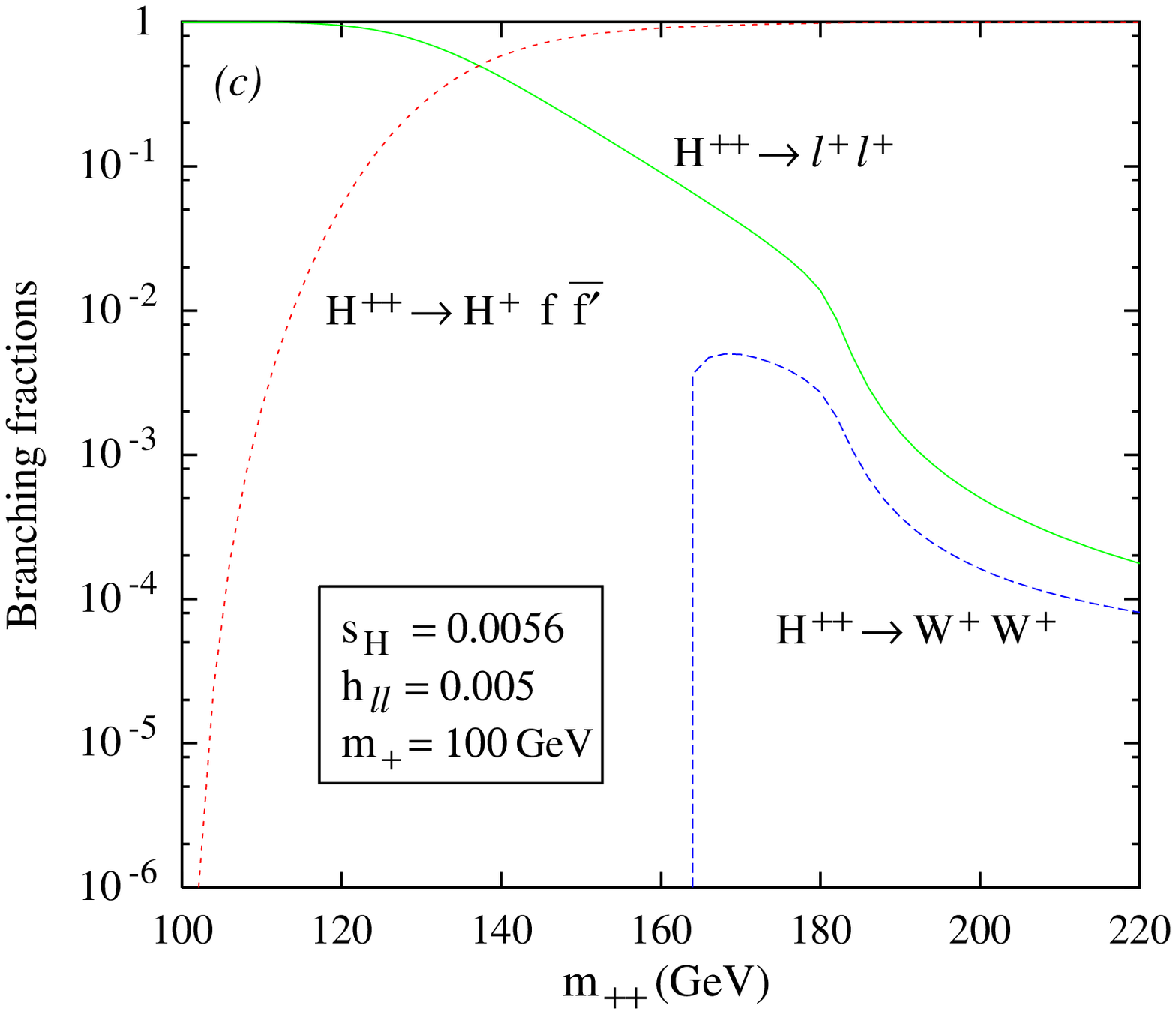,width=8.5cm,height=7cm,angle=0}
            \psfig{figure=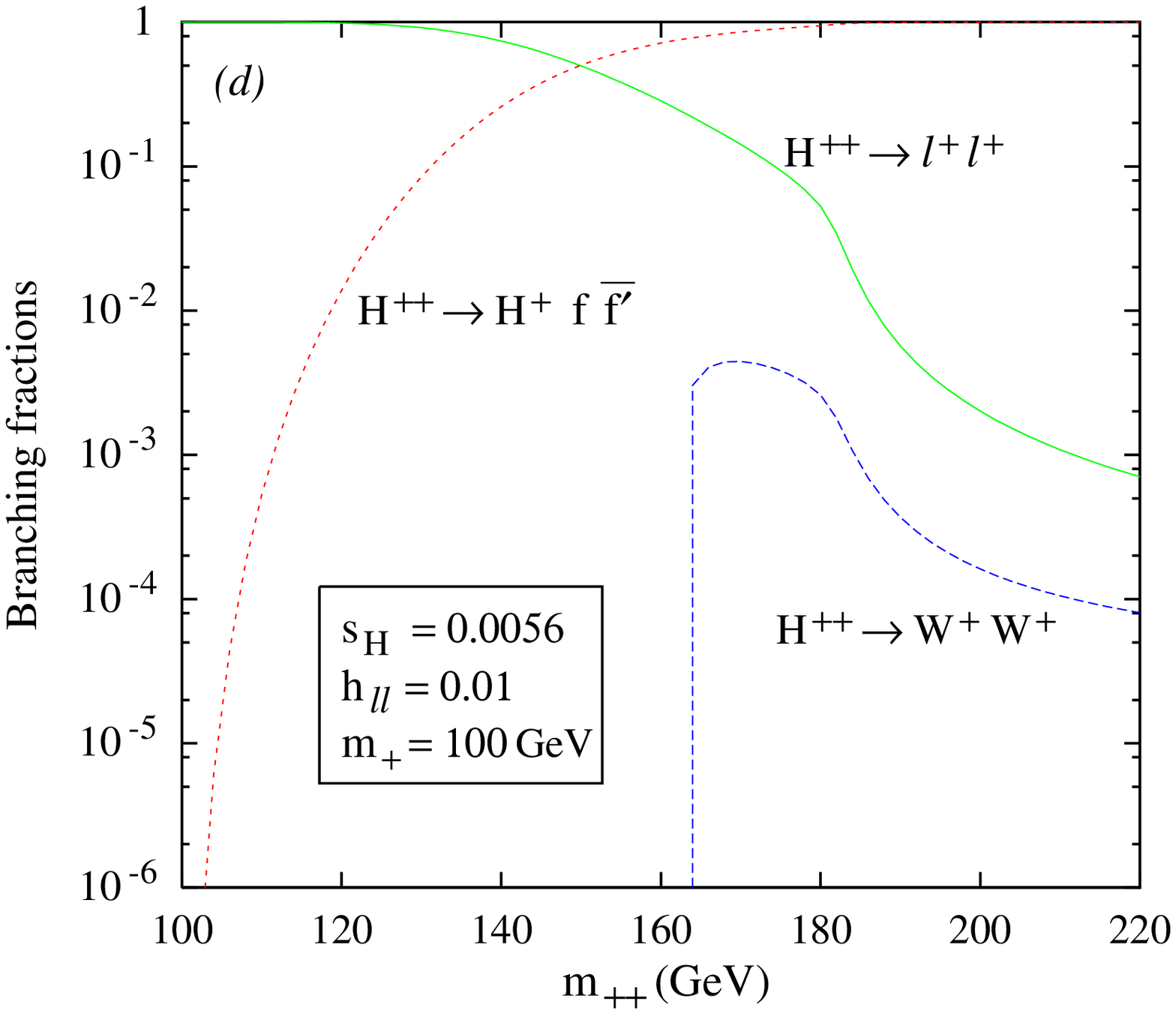,width=8.5cm,height=7cm,angle=0}
                   }
        \caption[sm]{\em The branching ratios for the different decay modes
of a doubly charged Higgs, for various combinations of parameters}
        \label{fig:branching}                                                   
\end{figure}

The above observations, however,  have to be balanced against the fact that
process (3) is in fact driven by the SU(2) gauge coupling, and is {\it not}
suppressed by any small model parameter. Consequently, this mode and the
resulting final states are definitely expected to dominate the signals of
a doubly charged Higgs over a substantial region of the parameter space.
This dominaton is of course more for cases where $|m_{++} - m_{+}|$ is
large enough for the two-body decay to be allowed. However, even when it is
not so, the decay $H^{++} \longrightarrow H^{+} W^*$, with the virtual W    
decaying fermionically, is a strong  competitor against the like-sign
dilepton channel, and  can even  dominate over it in certain regons. 
This fact has been demonstrated in 
Figs \ref{fig:branching}($a$--$d$), where we plot the branching 
fractions of an $H^{++}$ into the various modes. The mass of the singly
charged Higgs has been fixed at 100 GeV in all the figures.

In all of these figures, the mode driven by gauge coupling is
unmistakably dominant once the two-body decay involving a real W is allowed.
Furthermore, especially when $h_{ll}$ is on the smaller side, the 
three-body decay rate also exceeds that for the $ll$ mode over a 
substantial region of the parameter space. 
The various combinations of
parameters adequately demonstate that for a phenomenological study of the 
signals of $H^{++}$, it is extremely important to give due weightage to
channel (3). 

The final signals should depend in such a case on the various channels
of $H^+$-decay, which can be two-fermion modes (driven by $h_{ll}$ or
$s_H$ and suppressed by $m_{f}/m_W$), or four-fermion final states which are
controlled by $m_H$ but are not suppressed by any light fermion mass. 
The relative importance of these different modes have already been
discussed in the literature \cite{us1,us2, apostolos}.

It has been assumed in the entire discussion above that the $H^{++}$ is
heavier than the $H^{+}$. However, in a phenomenological analysis,
one should also take into account when it is the other way around. In such a
case, the hitherto dominant decay becomes kinematically disallowed, and  
 the channel $H^{++} \longrightarrow l^+ l^+$ competes with 
the $WW$ mode. In regions where the latter are kinematically suppressed,
the former is more or
less uniformly dominant, and the like-sign dilepton pairs constitute the
most important signal. However, there is a rather interesting possibility where
the dileptonic mode is the only viable one, but $h_{\tau\tau}$ is so small
that the decay length is very large. (For example, if the tau neutrino has
to be stable, then  the bound on $h_{\tau\tau}$ given in section 2 
gets tightened by some six orders of magnitude for a given $s_H$.) With, say,
$h_{\tau\tau}$ = O($10^{-9}$), and $m_{++}$ = $150$ GeV, the decay length
can be as large as 20 meters. This kind of a situation can give rise to
signals in the form of thickly ionised tracks (corresponding to heavy
long-lived doubly charged scalars) followed by `delayed' events, where the
ditaus may be registered as far out as the muon detectors.

\section{CONCLUSIONS}

We have pointed out that the $\gamma\gamma$ option in a  high energy
electron-positron collider can be a copious source of doubly charged 
scalars. Our analysis also shows that the production rates are experimentally
significant for a rather large range of $m_{H^{++}}$ in a 500 GeV linear
$e^{+}e^{-}$ collider, with some interesting dependence on the polarisation 
choice, which can be utilised in increasing the signal strength and 
hence the range of $m_{++}$ that can be probed.

As for the decay of the doubly charged scalars, we have demonstrated that
the like-sign dilepton mode is not always the most dominant one, especially 
when a singly singly charged scalar is there for the former to decay into.
In such cases, the dominant decay channel is mostly the decay into the 
singly charged scalar together with a real or a virtual W. This should lead to
signals with a large multiplicity of fermions in the final state. It is,
therefore, extremely important to consider such signals and find ways to
eliminate the corresponding backgrounds, if one is interested in 
observing doubly charged Higgs bosons.

{\bf Acknowledgements}: We thank the participants and organisers of the
Fourth Workshop on High Energy Particle Phenomenology, Calcutta, 
(funded by the Department of Science and Technology, Government of India), 
where this project was started. S.C. is grateful to
Amitava Datta and Amitava Raychaudhuri for giving him access to their
computing facilities.


\begin{thebibliography}{99}
\bibitem{hhunter} For a  general review see, J.F. Gunion, H.E. Haber, G.L. Kane
and S. Dawson, {\it The Higgs Hunters Guide}, Addison-Weseley, Reading, MA,
1990 and references therein. 
\bibitem{moh} R. N. Mohapatra and G. Senjanivic, Phys. Rev. Lett. {\bf 44}
(1980) 912, Phys. Rev. {\bf D 23} (1981) 165.
\bibitem{neutr} H. M. Georgi, S. L. Glashow and S. Nussinov, Nucl. Phys.
{\bf B 193} (1981) 297;  P.B. Pal, Phys. Rev. {\bf D 30}(1984) 2100;
L.I. Li, Y. Liu and L. Wolfenstein, Phys. Lett. {\bf B 159} (1985) 45.
\bibitem{gunion} For an early discussion of the phenomenology of models 
with Higgs triplets, see for example, J.F. Gunion, R. Vega and J. Wudka,
Phys. Rev. {\bf D 42} (1990) 1673.
\bibitem{us1} B. Mukhopadhyaya, Phys. Lett. {\bf B252} (1990) 123;
R.M. Godbole, B. Mukhopadhyaya and M. Nowakowski, Phys.
Lett. {\bf B 352} (1995) 388.
\bibitem{us2} D.K. Ghosh, R.M. Godbole and B. Mukhopadhyaya,
Phys. Rev. {\bf D 55} (1997) 3150.
\bibitem{apostolos} K. Cheung, R.J.N. Phillips  and A. Pilaftisis, 
Phys. Rev. {\bf D 51} (1995) 4731.
\bibitem{swartz} M. L. Swartz, Phys. Rev. {\bf D 40} (1989) 1521.
\bibitem{picciotto} C. Picciotto, Phys. Rev. {\bf D 56} (1997) 1612.
\bibitem{raidal} G. Barenboim, K. Huitu, J. Maalampi and M. Raidal,
Phys. Lett. {\bf B 394} (1997) 132.
\bibitem{lusignoli} M. Lusignoli and S. Petrarca, Phys. Lett. {\bf 226}
(1989) 397.
\bibitem{cuypers} F. Cuypers and M. Raidal, Nucl. Phys.. {\bf B 501}
(1997) 3.
\bibitem{loomis} J.F. Gunion, C. Loomis and K.T. Pitts, {\bf hep-ph/9610327}, 
To appear in {\it Proceedings of the 1996 DPF/DPB Summer study on 
		  New Directions for High energy Physics} .
\bibitem{raidal2} K. Huitu, J. Maalampi, A. Pietila and M. Raidal,
Nucl. Phys. {\bf B 487} (1997) 27. 
\bibitem{ackr} A. G. Akeroyd, hep-ph/9803324.
\bibitem{fawzi} For a review see , for example, M. Baillargeon, G. Belanger and
F. Boudjema,  {\it Proceedings of Two Photon Physics from Daphane to LEP 200
and Beyond Paris, Feb. 1994,  267-308}.
\bibitem{slac} D. Burke et al, Phys. Rev. Lett. {\bf 79} (1987) 1626.
\bibitem{telnov} I.F. Ginzburg et al, Nucl. Inst. Meth., {\bf205} (1983) 47;
I.F. Ginzburg et al, Nucl. Inst. Meth., {\bf 219} (1984);
V. Telnov, Nucl. Inst. Meth., {\bf A 355}(1995) 3.
\bibitem{pdg} {\it Review of Particle Properties}, Particle Data Group, 
Phys. Rev. {\bf D54}, (1996) 1.
\bibitem{manchek} H. Georgi and M. Machacek, Nucl. Phys. {\bf B 329}
(1985) 463.
\bibitem{golden} S. Chanowitz and M. Golden, Phys. Lett. {\bf B 165} (1985)
105.

\end{thebibliography}
\end{document}